\documentclass[10pt,aps,prl,letterpaper,twocolumn,superscriptaddress,showkeys]{revtex4-2}
\usepackage{color}
\usepackage{graphicx}
\usepackage{amsmath}
\usepackage{amssymb,amsthm}
\usepackage{hyperref}
\usepackage{mathtools}

\graphicspath{{pict/}{}}

\usepackage[applemac]{inputenc}

\usepackage{pgf,tikz}
\usepackage{mathrsfs}
\usepackage{bm}
\usepackage{array}
\usepackage{blkarray}
\usepackage{xcolor}

\usepackage{hyperref}
\hypersetup{colorlinks=true,linkcolor=blue,citecolor=blue,urlcolor=blue}
\global\long\def\ket#1{\left|#1\right\rangle }
\global\long\def\bra#1{\left\langle #1\right|}

\begin{document}
	

\title{Alice and Bob through a quantum mirror}


\author{M. Uria}
\email{maruria@udec.cl}
\affiliation{Facultad de Ciencias F\'isicas y Matem\'aticas, Departamento de F\'isica, Universidad de Concepci\'on, Concepci\'on, Chile}

\author{C. Hermann-Avigliano}
\affiliation{ANID - Millenium Science Iniciative Program - Millenium Institute for Research in Optics}
\affiliation{Departamento de F\'isica, Facultad de Ciencias F\'isicas y Matem\'aticas, Universidad de Chile, Santiago, Chile}

\author{P. Solano}
\affiliation{Facultad de Ciencias F\'isicas y Matem\'aticas, Departamento de F\'isica, Universidad de Concepci\'on, Concepci\'on, Chile}

\author{A. Delgado}
\email{aldelgado@udec.cl}
\affiliation{ANID - Millenium Science Iniciative Program - Millenium Institute for Research in Optics}
\affiliation{Facultad de Ciencias F\'isicas y Matem\'aticas, Departamento de F\'isica, Universidad de Concepci\'on, Concepci\'on, Chile}

\begin{abstract}
A quantum mirror is a device whose optical response, that is, transmission and reflection, can be controlled by a single qubit. Here, we propose the use of quantum mirrors as nodes in quantum networks. Propagating coherent states mediate the interaction between the control qubits of each quantum mirror. This allows implementing quantum teleportation, quantum state transfer, and entanglement swapping with success probability and average fidelity exponentially approaching unity as the average photon number increases. Furthermore, we show that quantum teleportation exhibits robustness against known sources of error, such as optical path phase difference, photon loss, and reduced quantum mirror reflectivity, presenting a promising alternative towards long-distance quantum communication. 
\end{abstract}


\maketitle


\textit{Introduction.---}A quantum network (QN) \cite {Bouwmeester,Cirac} is a collection of quantum nodes linked through quantum and classical channels. Each node can generate, store, and measure quantum states. The quantum channels enable the transmission of quantum states between subsets of nodes and offer an exponential increase of the space state dimension over classical connectivity. The QNs are complemented with quantum repeaters \cite{Azuma}, key components in the long-distance transmission of quantum states between nodes. Supplementing the nodes of a QN with quantum processors leads to the notion of a quantum internet \cite{Kimble,Wehner}. Applications of QNs include quantum key distribution \cite{Zhang}, secure access to remote quantum processors \cite{Broadbent}, distributed quantum computing \cite{Barral}, clock sincronization \cite{Komar}, quantum sensors \cite{Guo}, and increasing the baseline of telescopes \cite{Gottesman}.

The transmission of quantum states between nodes is implemented by quantum teleportation \cite{Bennett_1993} (QT) or quantum state transfer \cite{Cirac} (QST) and entanglement swapping \cite{Pan_1998}(ES) for quantum repeaters \cite{Briegel_1998} (QR). At each node, quantum states are usually encoded in matter qubits, and photons are used to generate entangled states between nodes  and to directly transmit quantum states between nodes.  QT has been demonstrated in a wide variety of quantum systems \cite{Pirandola_2015,Hu_2023} such as photonic qubits \cite{Bouwmeester_1997,Boschi_1998,Marcikic_2003,Ursin_2004,Yin_2012}, optical modes \cite{Furusawa_1998,Yonezawa_2007,Takeda_2013,Liu_2020}, atomic ensembles \cite{Sherson_2006,
Krauter_2013,Chen_2008}, trapped atoms \cite{Barrett_2004,Riebe_2004,Riebe_2007,Olmschenk_2009}, solid state systems \cite{Gao_2013,Bussieres_2014,Steffen_2013,Pfaff_2014,Llewellyn_2020}, nuclear magnetic resonance \cite{Nielsen_1998}, and quantum dots \cite{Strobel_2025}, with remarkable implementation of Earth-to-satellite teleportation \cite{Ren_2017}.  However,  the lack of a 100\% \ efficient Bell measurement \cite{Weinfurter_1994,Braunstein_1995,Calsamiglia_2001} leads to probabilistic realizations of QT and lower average fidelities. Teleportation fidelity can exceed the classical limit of $2/3$ \cite{Massar_1995} when postselecting for successful measurement, at the expense of reducing the efficiency of QT.  Additionally, photon losses reduce the teleportation distance when using single-photon quantum channels. 

These roadblocks motivate the search for new platforms to implement higher-performance QNs. Recently, it has been shown theoretically \cite{Bekenstein_2020} and experimentally \cite{Srakaew_2023} that the optical response of a quantum metasurface, that is, a subwavelength two-dimensional atomic array, can be coherently controlled by the quantum state of a single atom. In particular, if this control atom is in the ground state (excited), light is transmitted (reflected). Thus, a superposition of these states leads to a superposition of reflected and transmitted light, which has led to these quantum devices being described as \textit{quantum mirrors} (QMs).  This device generates entanglement between two propagating field modes and the control atom.

\begin{figure}
        
    \hspace{-0.5cm}\includegraphics[width=0.97\columnwidth]{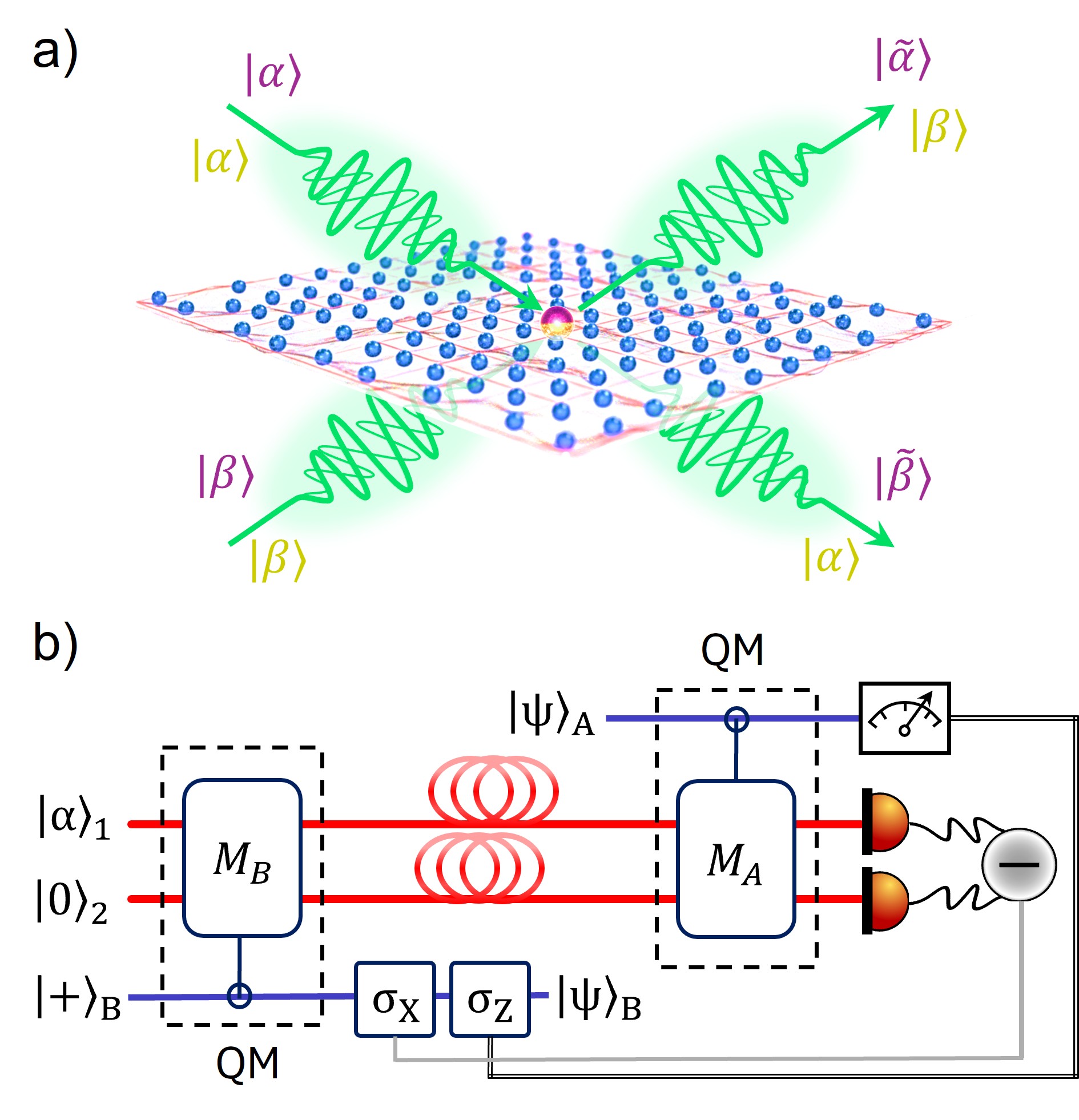}
    \caption{ (a) Schematic representation of a quantum mirror (QM) interacting with two optical modes. Yellow (purple) states denote transmission (reflection) conditioned on the control atom state $\ket{0}$ ($\ket{1}$). (b) Quantum circuit for the teleportation protocol between the control atoms of two QMs. Two optical modes (red) propagate from Bob to Alice, becoming entangled with Bob's control atom upon interaction. The modes then interact with Alice's QM, after which her system is measured. The measurement outcomes are classically communicated to Bob, who applies the appropriate operations to reconstruct the state initially prepared by Alice.}
    \label{fig:QM}
\end{figure}

In this article, we show that endowing each node of a QN with a QM enables high efficiency QT, QST, and ES, where qubits are encoded in the control atom of QMs. For example, in QT between two nodes, the control atom of the first QM is initialized in an equally weigthed superposition of ground and excited states and fed with a coherent state at each input port. The fields propagate toward the second QM, whose control atom is initialized in the unknown qubit state to be teleported. This control atom and the field modes are measured after interaction. Thereby, the unknown state is teleported to the control atom of the first QM, up to local Pauli operations. This teleportation scheme uses coherent states instead of single-photon states, making it well suited for long-distance demonstrations over free-space links or optical fibers. QT is carried out with a theoretical success probability and fidelity that asymptotically approach 1 as the average photon number of the coherent states increases. Furthermore, QT shows robustness against known sources of error, such as photon loss and phase mismatch, and against reduced reflectance of the QM. We show that similar protocols can lead to high-efficency QST and ES. These results indicate that QMs can be a well suited platform to implement QNs.

\textit{Quantum mirror.---}A quantum mirror implements a transformation that acts onto two distinguishable spatial modes of the quantized electromagnetic field conditional on the state of a control atom. This is given by \cite{Bekenstein_2020}
\begin{eqnarray}
{M}|0\rangle_c|\alpha\rangle_1|\beta\rangle_2&=&|0\rangle_c|\alpha\rangle_1|\beta\rangle_2,
\nonumber\\
{M}|1\rangle_c|\alpha\rangle_1|\beta\rangle_2&=&|1\rangle_c|\tilde\beta\rangle_1 |\tilde\alpha\rangle_2,
\label{M}
\end{eqnarray}
where the states $\{|0\rangle_c,|1\rangle_c\}$ are a basis of the control atom $c$, states $|\alpha\rangle_1$ and $|\beta\rangle_2$ describe coherent states with average photon numbers $|\alpha|^2$ and $|\beta|^2$ in modes 1 and 2 of the field, respectively, and $\tilde\beta=-\beta$ and $\tilde\alpha=-\alpha$ as shown in Fig.~\ref{fig:QM}(a).

The quantum mirror transformation leads to the perfect transmission of coherent states in modes 1 and 2 when the control atom is in the $|0\rangle$ state. Perfect reflection is obtained when the control atom is in the $|1\rangle$ state, in which case the modes are swapped and both coherent states gain a phase of $\pi$.

\textit{Quantum Teleportation.---}Two distant nodes, Alice and Bob, are provided with QMs so that they can implement the transformation in Eq.~(\ref{M}). The teleportation takes place from Alice to Bob. First, Bob applies the transformation on modes 1 and 2 initialized in the arbitrary states $|\alpha\rangle_1$ and $|\beta\rangle_2$, respectively, with the control atom in the state $(|0\rangle_{B}+|1\rangle_{B})/\sqrt{2}$. Thereby, Bob generates the state
\begin{equation}
\frac{1}{\sqrt{2}}(|0\rangle_{B}|\alpha\rangle_1|\beta\rangle_2+
|1\rangle_{B}|\tilde\beta\rangle_1|\tilde\alpha\rangle_2).
\label{Channel}
\end{equation}
Photons in modes 1 and 2 propagate towards Alice's QM, whose control atom is prepared in the unknown state to be teleported $|\psi\rangle_A=a|0\rangle_A+b|1\rangle_A$. Thus, after the QM transformation on Alice's side, the control atoms and the modes are described by the state
\begin{eqnarray}
&&\frac{1}{2}|+\rangle_A(|\psi\rangle_B|\alpha\rangle_1|\beta\rangle_2
+\sigma_x^B|\psi\rangle_B|\tilde\beta\rangle_1|\tilde\alpha\rangle_2)+
\nonumber\\
&&\frac{1}{2}|-\rangle_A(\sigma_z^B|\psi\rangle_B|\alpha\rangle_1|\beta\rangle_2
-\sigma_z^B\sigma_x^B|\psi\rangle_B|\tilde\beta\rangle_1|\tilde\alpha\rangle_2),
\label{ProtoTeleportation}
\end{eqnarray}
where $|\pm\rangle_A=(|0\rangle_A\pm|1\rangle_A)/\sqrt{2}$, $|\psi\rangle_B=a|0\rangle_B+b|1\rangle_B$ and $\{\sigma_x^B,\sigma_z^B\}$ are Pauli operators acting on Bob's control atom. 

\begin{figure}[t]
    \centering
    \includegraphics[width=0.95\linewidth]{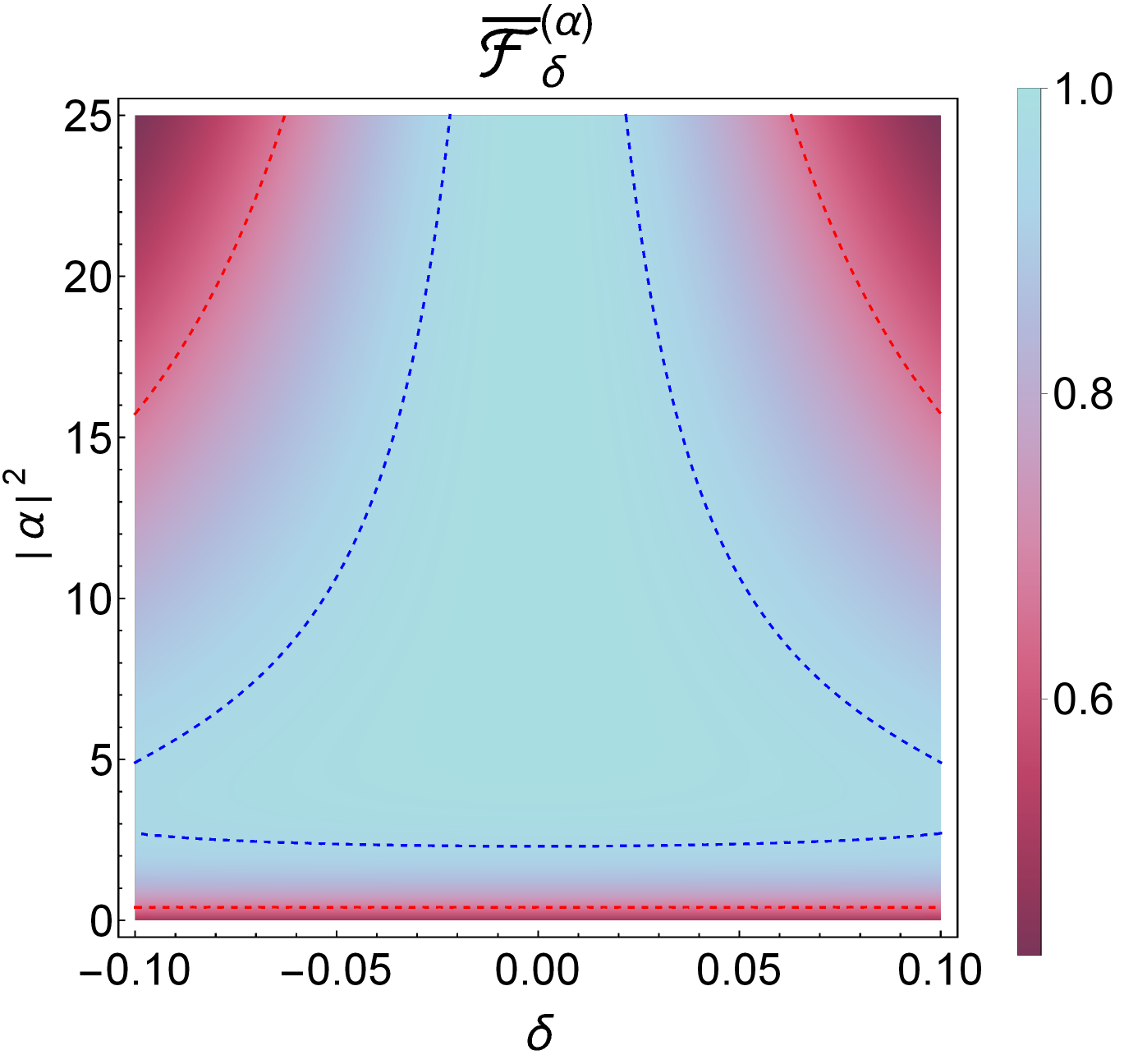}
    \caption{Average teleportation fidelity $\bar{\mathcal{F}}_{\delta}^{(\alpha)}$ as a function of $\delta$, the phase difference between optical  modes, and $|\alpha|^2$, the average number of photons in the propagating field. Area enclosed by the dashed blue lines describe a parameter region where $\bar{\mathcal{F}}_{\delta}^{(\alpha)}$ is greater than 0.95. Area enclosed by the dashed red lines describe a parameter region where $\bar{\mathcal{F}}_{\delta}^{(\alpha)}$ is greater than 0.66, the classical threshold.}
    \label{fig:delta}
\end{figure}

The state in Eq.\thinspace(\ref{ProtoTeleportation}) already exhibits the main features of QT. Each of the four terms of this state contains the state to be teleported in Bob's control atom, up to local Pauli operations. Furthermore, each of these terms has a unique combination of states for Alice's control atom and field modes. For example, after projecting Alice's control atom onto state $|+\rangle_A$, the teleportation procedure can be completed if Alice can reliably distinguish states $|\alpha\rangle_1|\beta\rangle_2$ and $|\tilde\beta\rangle_1|\tilde\alpha\rangle_2$. However, the latter are not necessarily orthogonal and therefore cannot be distinguished with certainty, a key factor that would decrease the fidelity of teleportation. A simple scenario can be obtained by assuming $\beta=0$ as shown in Fig.~\ref{fig:QM}(b), in which case Alice must discriminate between states $|\alpha\rangle_1|0\rangle_2$ and $|0\rangle_1|\tilde\alpha\rangle_2$. This can be done by placing photodetectors at each output port of Alice's quantum mirror. Since these lead to detection at different output ports, one detection allows the state to be uniquely identified. Depending on the measurement results, Pauli operators are applied by Bob to his control atom as indicated in Eq.~\eqref{ProtoTeleportation}. In the event of no photo-detections at both output ports, the state received by Bob is $|+\rangle_B$, which has no information about $\ket{\psi}$. The total failure probability $P_f$ due to this event is
\begin{equation}
P_f=e^{-|\alpha|^2},    
\end{equation}
which vanishes exponentially with the average number of photons. Thus, with success probability $P_s=1-e^{-|\alpha|^2}$ the states are perfectly teleported.

The performance of the teleportation scheme is evaluated using average fidelity $\bar{\mathcal{F}}$ as a metric. This is the fidelity $\mathcal{F}=|\langle\psi|\tilde\psi\rangle|^2$ averaged over the effectively teleported states $|\tilde\psi\rangle$ and over all possible states $|\psi\rangle$ that can be teleported, the latter selected according to the uniform Haar distribution. The average fidelity considers all measurement results, successful or not. For our teleportation scheme, we obtain
\begin{equation}
\bar{\mathcal{F}}^{(\alpha)}=1-\frac{1}{2}e^{-|\alpha|^2}.
\end{equation}
Therefore, the average fidelity approaches 1 exponentially as the average number of photons increases. $\bar{\mathcal{F}}^{(\alpha)}\ge0.99$ is reached with $|\alpha|^2\ge4$, in which case $P_s\ge0.98$. More noticeable, for $|\alpha|^2>\ln(3/2)\approx 0.40$ the average fidelity is already greater than the classical limit of $2/3$.  

\begin{figure}[t]
    \centering
    \hspace{-0.25cm}\includegraphics[width=\linewidth]{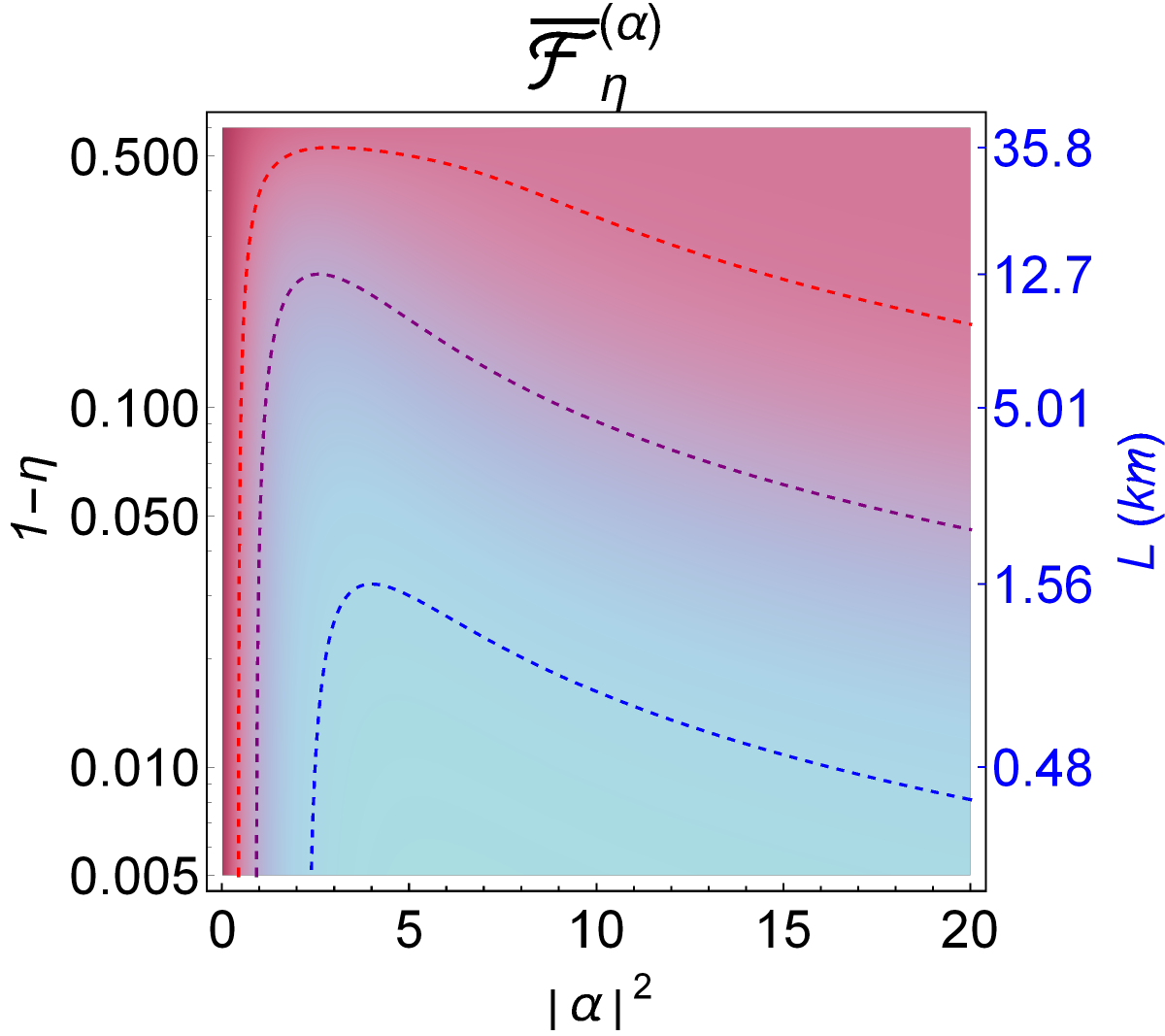}
    \caption{Average teleportation fidelity $\bar{\mathcal{F}}_{\eta}^{(\alpha)}$ as a function of $\eta$, the channel's transmission efficiency, and $|\alpha|^2$, the average number of photons in the propagating field. The area below the dashed line describes a parameter region with average fidelity greater than: 0.95 (blue), 0.80 (purple) and 0.66 (red). The right-hand blue labels indicate the corresponding fiber lengths in kilometers (km) for select values of $\eta$, specially corresponding to the maximum average fidelities achieved along the dashed lines, considering a realistic and state-of-the-art loss rate of  $\gamma=6.3\text{kHz}$. Color scale as in Fig.~\ref{fig:delta}.}
    \label{fig:decoherence}
\end{figure}

\textit{Quantum State Transfer.---} For this purpose \cite{Cirac}, Alice initializes its control atom in the state to be transferred, $|\psi\rangle_A$, and feeds the QM with the two-mode state $|\alpha\rangle_1|0\rangle_2$, yielding
\begin{equation}
a|0\rangle_A|\alpha\rangle_1|0\rangle_2+b|1\rangle_A|0\rangle_1|\tilde\alpha\rangle_2.
\end{equation}
Alice continues by measuring the control atom on the basis $|\pm\rangle_A$, producing the locally equivalent states
\begin{equation}
\frac{1}{\sqrt{P_\pm}}(a|\alpha\rangle_1|0\rangle_2\pm b|0\rangle_1|\tilde\alpha\rangle_2),
\end{equation}
with probability $P_\pm= 1/2 \, \pm \,\text{Re}[a^*b]e^{-\frac{1}{2}|\alpha|^2}$, respectively. Fields in modes 1 and 2 are transmitted towards Bob's QM, which is initialized in the state $|+\rangle_B$, generating the states 
\begin{equation}
    \frac{1}{\sqrt{2P_+}}(|\psi\rangle_B|\alpha\rangle_1|0\rangle_2+\sigma_x^B|\psi\rangle_B|0\rangle_1|\tilde\alpha\rangle_2)
\end{equation}
or
\begin{equation}
    \frac{1}{\sqrt{2P_-}}\sigma_z^B(|\psi\rangle_B|\alpha\rangle_1|0\rangle_2-\sigma_x^B|\psi\rangle_B|0\rangle_1|\tilde\alpha\rangle_2),
\end{equation}
depending on the measurement of Alice's QM. Finally, depending on the measurement results in the field modes, Bob applies Pauli operators to his control atom to fully transfer $\ket{\psi}$. Note that QST does not require entanglement between nodes, nor between node and field modes during propagation. The implementation of quantum state transfer has a total success probability of $1-e^{-|\alpha|^2}$ and, if the discrimination state fails, the transferred state is $|+\rangle_B$. Thereby, QT and QST via QMs lead to the same average fidelity. 

\textit{Entanglement swapping.---}Now suppose that light cannot be transmitted directly from Bob to Alice. In this case, a third node, Charlie, also equipped with a QM, is added between Alice and Bob. The latter prepares the state Eq.~(\ref{Channel}) with $\beta=0$, that is,
\begin{equation}
\frac{1}{\sqrt{2}}(|0\rangle_{B}|\alpha'\rangle_1|0\rangle_2+
|1\rangle_{B}|0\rangle_1|\tilde\alpha'\rangle_2),
\label{Channel2}
\end{equation}
and sends the photons on modes 1 and 2 toward Charlie, who initializes the control atom of his QM in the state $(|0\rangle_C+|1\rangle_C)/\sqrt{2}$. The QM transformation on Charlie's side generates the following state
\begin{equation}
\frac{1}{\sqrt{2}}|\phi\rangle_{CB}|\alpha'\rangle_1|0\rangle_2
+\frac{1}{\sqrt{2}}|\psi\rangle_{CB}|0\rangle_1|\tilde\alpha'\rangle_2,
\end{equation}
where $|\phi\rangle_{CB}=(|0\rangle_C|0\rangle_B+|1\rangle_C|1\rangle_B)/\sqrt{2}$ and $|\psi\rangle_{CB}=\sigma_x^{C}|\phi\rangle_{CB}$. These Bell states are generated whenever Charlie successfully distinguishes the states $|\alpha'\rangle_1|0\rangle_2$ and $|0\rangle_1|\tilde\alpha'\rangle_2$. Depending on the field measurement results, Charlie applies $\sigma_x^{C}$ to his control atom, so he always prepares the state $\ket{\phi}_{CB}$. Next, Charlie feeds his QM with a new state $|\alpha\rangle_1|0\rangle_2$ that after the QM transformation leads to  
\begin{eqnarray}
&&\frac{1}{\sqrt{2}}|+\rangle_C\frac{1}{\sqrt{2}}(|0\rangle_B|\alpha\rangle_1|0\rangle_2+|1\rangle_B|0\rangle_1|\tilde\alpha\rangle_2)
\nonumber\\
&&+\frac{1}{\sqrt{2}}|-\rangle_C\frac{1}{\sqrt{2}}(|0\rangle_B|\alpha\rangle_1|0\rangle_2-|1\rangle_B|0\rangle_1|\tilde\alpha\rangle_2).
\end{eqnarray}
After measuring Charlie's control atom on the basis $\{|\pm\rangle_c\}$, two locally equivalent states under $\sigma_z^B$ are generated. The fields in modes 1 and 2 can now be transmitted from Charlie to Alice, which can be used as a resource for implementing the teleportation between Alice and Bob. The total success probability of this process is $1-e^{-|\alpha'|^2}$. The concatenation of $N$ of these processes has a total success probability of $(1-e^{-|\alpha'|^2})^N$, which can be approximated as $1-Ne^{-|\alpha'|^2}$ for $|\alpha'|^2$ large. 

\textit{Performance analysis of quantum teleportation.---} Next, we will assume that QT takes place without using any correction stage. The first source of error occurs when considering an optical path difference between optical modes, leading to a phase difference $\delta$. The average fidelity becomes
\begin{equation}
\bar{\mathcal{F}}_{\delta}^{(\alpha)}=\bar{\mathcal{F}}^{(\alpha)}-\frac{1-e^{-|\alpha| ^2 (1-\cos\delta )} \cos \left(|\alpha| ^2 \sin\delta \right)}{3}.
\end{equation}
for $\delta$ small enough. As can be seen, the average fidelity is reduced by the phase difference with respect to the case $\delta=0$. However, for small values of the average photon number, the decrease still exceeds the classical fidelity limit of $2/3$. Figure~\ref{fig:delta} shows a density plot of $\bar{\mathcal{F}}_{\delta}^{(\alpha)}$ as a function of $|\alpha|^2$ and $\delta$ for two level curves: red dashed line (blue) for fidelity values equal to $0.66$ ($0.95$). According to this figure, it is possible to teleport with high fidelity over a wide range of $\delta$ and $|\alpha|^2$ values, as long as the product $(|\alpha|\delta)^2$ is small. 

Photon loss during propagation through optical fibers represents another important error source, which can be accurately modeled as an amplitude damping channel \cite{Liu_2004}. Considering two distinct optical modes, labeled 0 and 1, each  independently interacting with a common zero-temperature environment, the time evolution of the initial joint state $\rho(0)$ is described by the Kraus map $\rho(t)=\sum_{m,k}(A_m^{(0)}\otimes A_k^{(1)})\rho(0)(A_m^{\dagger(0)}\otimes A_k^{\dagger(1)})$ where the Kraus operator $A_k(t)=\sum_{i=k}\sqrt{\binom{i}{k}}\eta_t^{(i-k)/2}(1-\eta_t)^{k/2}\ket{i-k}\bra{i}$  represents the loss of $k$ photons from a single mode. Here, $\eta_t=e^{-\gamma t}$ is the channel's transmission efficiency at time $t$, with the damping rate $\gamma$. Under this loss model, the average fidelity becomes
\begin{eqnarray}
\bar{\mathcal{F}}_{\eta}^{(\alpha)}=\bar{\mathcal{F}}^{(\alpha)}-\frac{1}{6}(1-e^{-|\alpha|^2(1-\eta)})(2+e^{-|\alpha|^2\eta}).
\end{eqnarray}

Figure~\ref{fig:decoherence} shows a density plot of the average fidelity of teleportation $\bar{\mathcal{F}}_{\eta}^{(\alpha)}$ as a function of the channel transmission efficiency $\eta$, the average number of photons $|\alpha|^2$, and the propagation distance $L$ for a realistic and state-of-the-art loss rate of $\gamma = 6.3  \text{kHz}$ \cite{Petrovich_2025}, which corresponds to 50$\%$ intensity loss over a distance of 32.96 km. The three level curves $\bar{\mathcal{F}}_{\eta}^{(\alpha)}=0.95, 0.80,$ and $0.66$ are indicated by blue, purple, and red dashed lines, respectively. As is apparent from this figure, for a fixed value of $\bar{\mathcal{F}}_{\eta}^{(\alpha)}$, the maximal teleportation distance is achieved for small values of the average number of photons, and high fidelity teleportation is still achievable for moderate losses. The classical threshold for teleportation is achieved close to 36 km. An average photon number $|\alpha|^2=4$  yields an average fidelity of $\bar{\mathcal{F}}_{\eta}^{(\alpha)}=0.95$ for a teleportation distance of 1.6 km.

Another important source of error arises from an imperfect realization of QMs. These manifest as errors only upon reflection, leading to a non-vanishing transmitivity. Taking into account this error in both QMs, the average fidelity becomes
\begin{equation}
\begin{aligned}
\bar{\mathcal{F}}^{(\alpha)}_{r_{1,2},t_{1,2}} =
\frac{1}{12}\Big[
6 - 5e^{-|\alpha|^2}
- e^{-|\alpha|^2 |r_1 r_2+t_1 t_2|^2}
+ e^{-|\alpha|^2 |t_1|^2} \\
+ e^{-|\alpha|^2 |t_2|^2} 
+ 2 e^{-|\alpha|^2 (1-\mathrm{Re}[r_1 r_2^*])}
\cos\!\big(|\alpha|^2\,\mathrm{Im}[r_1 r_2^*]\big) \\
+ 2 e^{-|\alpha|^2 (1-\mathrm{Re}[r_1 r_2+t_1t_2])}
\cos\!\big(|\alpha|^2\,\mathrm{Im}[r_1 r_2+t_1 t_2]\big)
\Big],
\end{aligned}
\end{equation}
\begin{figure*}[t]
    \centering
    \includegraphics[width=0.9\linewidth]{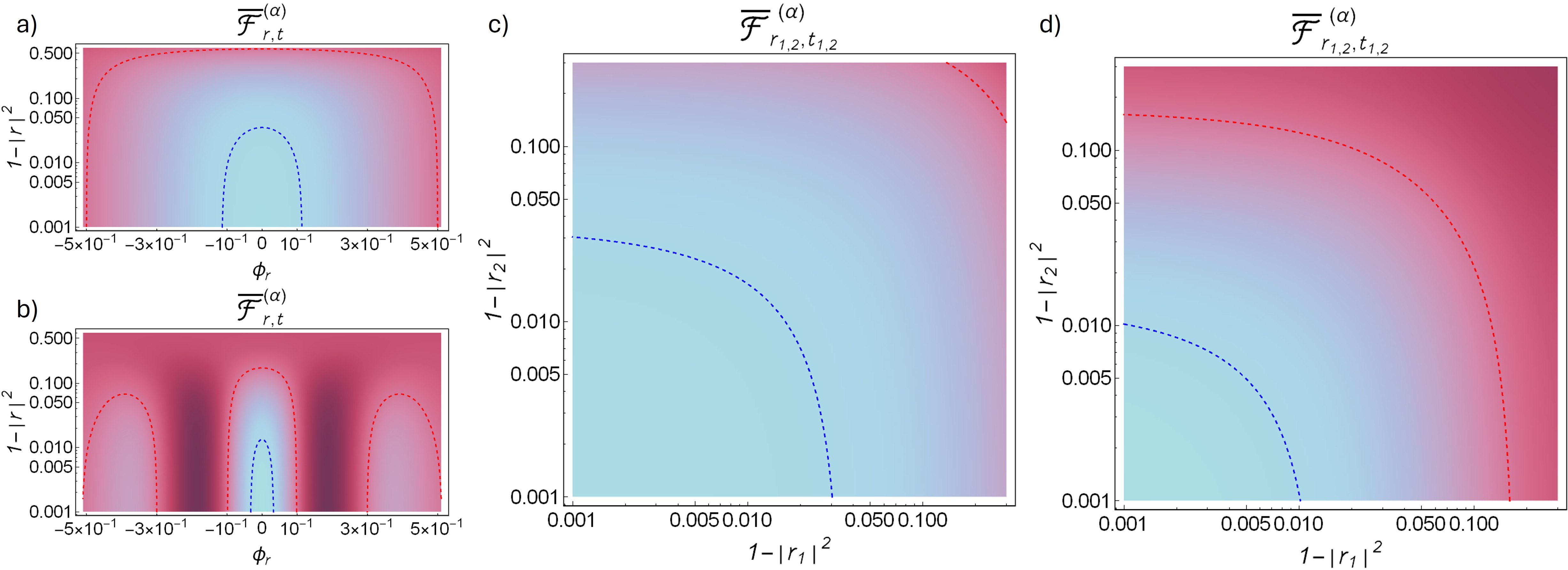}
    \caption{Average fidelities for errors in the amplitude $r$ and phase $\phi_r$ of reflection coefficients, presented for two different input field amplitudes: $\alpha=1.75$ and $\alpha=4$. Panels (a) and (b) depict the impact of errors in the Bob's quantum mirror. Panels (c) and (d) illustrate the effect of errors, assuming no phase errors, for both QMs.}
    \label{fig:mirrors_error}
\end{figure*}
where $r_1=\cos(\pi+\epsilon_1)e^{i\phi_1}$ and $t_1=\sin(\pi+\epsilon_1)e^{i(\phi_1+\pi/2)}$ are the coefficients describing the action of Bob's QM and $r_2=\cos(\pi+\epsilon_2)e^{i\phi_2}$, $t_2=\sin(\pi+\epsilon_2)e^{i(\phi_2+\pi/2)}$ are analogous coefficients for Alice's QM. The parameter pairs $(\epsilon_1,\phi_1)$ and $(\epsilon_2,\phi_2)$ characterize the deviations from the ideal operation of each QM. Figure~\ref{fig:mirrors_error}(a) shows a density plot of the average fidelity, under the assumption that only the Bob's QM manifest errors, as a function of $1-|r|^2$ and the reflection phase $\phi_r$ for the case that both QMs have a non unitary reflection coefficient that allows a amount of transmitted light and for $|\alpha|^2=1.75$. For small values of the transmission coefficient and of the phase it is possible to attain an average fidelity above 0.95 (blue dashed line) and even when 50\% of light is transmitted the average fidelity can still be above the classical fidelity threshold of 0.66. A similar result is observed in Fig.~\ref{fig:mirrors_error}(b) in the case $|\alpha|^2=4$, although the corresponding regions of parameters become smaller. Figure~\ref{fig:mirrors_error}(c) shows a density plot for the average fidelity as a function of $1-|r_1|^2$ and $1-|r_2|^2$ in the absence of phase errors ($\phi_1=\phi_2=0$) for $|\alpha|^2=1.75$. For small values of the transmission coefficients, an area (bounded by the blue dashed line) of fidelity values greater than 0.95 is present, although higher values of the transmission coefficients still provide average fidelity values above the classical threshold. A similar behavior is depicted in Fig.~\ref{fig:mirrors_error}(d) for $|\alpha|^2=4$.

\textit{Discussion and Outlook.---}
    Although the first proof-of-principle implementations of a QM using two-dimensional atomic arrays reported a reflectivity of nearly 50$\%$ \cite{Srakaew_2023}, purpose-built platforms can lead to further improvement. Reflectivity can reach 100\% for a lattice spacing close to 0.8 times the resonant wavelength \cite{Shahmoon_2017}. This can be achieved with a red-detuned dipole trap created by counter-propagating beams at 1.6 times the resonance wavelength of the trapped atoms. Furthermore, selecting atomic species with optical transitions within the telecommunication band, such as $^{166}Er$ \cite{Grun_2024} and $^{171}Yb$ \cite{Covey_2019}, could minimize propagation losses, while an auxiliary off-resonant laser could be used to stabilize the phase of the optical communication channel.
    
    QMs can also be implemented on alternative configurations, such as waveguide QED \cite{Sinha_2025}, and  on alternative platforms, such as microwave fields interacting with superconducting qubits. These systems have demonstrated the ability to transmit or reflect photons conditioned on the state of a single qubit \cite{Mirhosseini_2019}, effectively creating a single-qubit QM. Because these microwave fields propagate through waveguides, a circulator is necessary to isolate incident from outgoing modes \cite{Fedorov_2024}. While this technology could lead to QT and QST protocols within superconducting quantum processors, microwave photons pose significant constraints for truly long-distance quantum communications.

As QMs constitute a source of entanglement, is not hard to envision other physical applications such as quantum estate estimation and quantum metrology. Furthermore, QT has been shown to be a primitive for quantum computing \cite{Gottesman_1999,Ishizaka_2008}, thus we can imagine an array of QMs communicating through coherent states to realize non-local quantum gates, naturally suitable for distributed quantum computing \cite{Barral}.


\textit{Summary.---}We have proposed the use of \textit{quantum mirrors} as nodes in quantum networks to implement high fidelity and high success probability quantum teleportation, quantum state transfer, and entanglement swapping. This is possible because of the capacity of \textit{quantum mirrors} to generate entangled states between the control atom and two spatial optical modes. The use of coherent states provides a simple and robust implementation with average fidelity and success probability exponentially approaching unity as the average photon number increases. We also show that \textit{quantum mirrors} provide robustness against error sources such as optical path phase difference, photon losses, and reduced reflectivity, enabling long-distance quantum communication.  Finally, we hope that the results presented here will contribute to motivating research on quantum-controlled metamaterials and their application to quantum information science.

\textit{Acknowledgments.---} The authors thank K. Sinha for helpful discussions. This work was supported in part by FONDECYT REGULAR No. 1230897 and No. 1240204, and Millennium Science Initiative Program ICN17$_-$012. M. U. was supported by ANID Doctoral Fellowship Grant No. 21232285.

\end{document}